# Antithesis of Object Orientation: Occurrence-Only Modeling Applied in Engineering and Medicine

Sabah Al-Fedaghi*

*Computer Engineering Department*
*Kuwait University*
*Kuwait*

salfedaghi@yahoo.com, sabah.alfedaghi@ku.edu.kw

***Abstract* – This paper has a dual character, combining a philosophical ontological exploration with a conceptual modeling approach in systems and software engineering. Such duality is already practiced in software engineering, in which the current dominant modeling thesis is object orientation. This work embraces an anti-thesis that centers solely on *the process* rather than emphasizing *the object*. The approach is called *occurrence-only* modeling, in which an *occurrence* means an *event* or *process* where a process is defined as an orchestrated net of events that form a semantical whole. In contrast to object orientation, in this occurrence-only modeling objects are nothing more than long events. We apply this paradigm to (1) a UML/BPMN inventory system in simulation engineering and (2) an event-based system that represents medical occurrences that occur on a timeline. The aim of such a venture is to enhance the field of conceptual modeling by adding yet a new alternative methodology and clarifying differences among approaches. Conceptual modeling's importance has been recognized in many research areas. An active research community in simulation engineering demonstrates the growing interest in conceptual modeling. In the clinical domains, temporal information elucidates the occurrence of medical events (e.g., visits, laboratory tests). These applications give an opportunity to propose a new approach that includes (a) a Stoic ontology that has two types of being, *existence* and *subsistence*; (b) Thinging machines that limit activities to five generic *actions;* and (c) Lupascian logic, which handles *negative events*. With such a study, we aim to substantiate the assertion that the "occurrence only" approach is a genuine philosophical base for conceptual modeling. The results in this paper seem to support such a claim.**

## I. Introduction

> The world is a network of events. Of happenings. Of processes. Of something that occurs. The things that are most "thinglike" are nothing more than long events. The hardest stone is in reality a complex vibration of quantum fields, a momentary interaction of forces, a process that for a brief moment manages to keep its shape, to hold itself in equilibrium before disintegrating again into dust. The world is not so much made of stones as of fleeting sounds, or of waves moving through the sea. [1]

> A war is not a thing, it's a sequence of events. A storm is not a thing, it's a collection of occurrences. A cloud above a mountain is not a thing, it is the condensation of humidity in the air that the wind blows over the mountain. A wave is not a thing, it is a movement of water, and the water that forms it is always different. [1]

This paper has a dual character, combining a philosophical ontological exploration with a conceptual modeling approach. On the one hand, the philosophical undertaking is an attempt to provide a representation of reality by modeling features of the world in a specific domain. On the other hand, to understand real "systems," i.e., case studies in sections 3 and 4, the developed representation has to be demonstrated in practical, reasonable size fields.

According to Shults [2], computer scientists have typically had little interest in philosophers' arguments about the nature of being(s) and non-being. In recent years, a growing number of scientists in the modeling community have explored various advances in their field that bear on philosophical issues related to ontology. Shults [2] claims that developments in computer modeling have the potential to contribute to what may be "the most significant change in western philosophy since the foundational work of Aristotle's teacher Plato in the 4th century BC."

In this context, we view a conceptual model as a depiction of reality built using diagrammatic construction that is oriented toward human communication. This diagrammatic orientation started with earlier examples, which include *states* in finite-state machines and *activities* in flowcharts, which lead to modeling languages such as SysML Object Process Methodology, UML and BPMN [3]. In most such modeling languages, it is claimed that reality conceptualization requires *objects* as a basic construct to express the system's structure and *processes* to grant the model understanding of the system's dynamic behavior [4][5]. This requires adopting such notions as classes and associations with attributes and operations, aggregation and generalization, and predefined relationships, claiming applicability in many real-world problems with ease of use.

---

*Retired June 2021, seconded fall semester 2021/2022

**This paper contests this approach, which is based on substance ("being" or "a basic entity"), wholly or partially, and challenges it as a fundamental paradigm.** Although such a dispute is not new (e.g., Whitehead process philosophy), the paper provides a more complete framework called *occurrence-only modeling*, with ontology and modeling language as an antithesis to object-oriented conceptual modeling, in which the thesis is substance-based ontology (e.g., Mario Bunge's ontology) and language such a UML is used to model real-world semantics.

The study presents a conceptualization based solely on *occurrences* (see Fig. 1). An occurrence is an *event* or *process*, and a process is defined as an orchestrated net of events that form a whole and emerge from these events. An event is a *subsisting region of potentiality* "activated" by *time*, as we will show in detail later in this paper. The approach name is quantified with "only" instead of "oriented" to highlight that it is not an *alignment toward*; rather, it is a total commitment to a technique that is based solely on occurrences.

The proposed occurrence-only modeling is specified as a high-level diagrammatic language using Stoic ontology [6], thinging machines (TM) [7], and Lupascian logic [8]. See Fig. 2 for important notions that we will discuss in this paper.

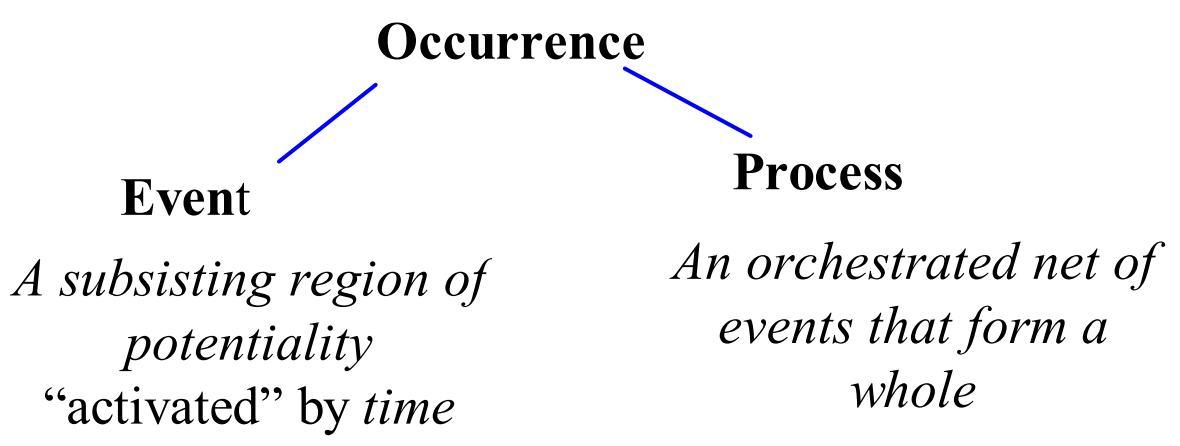

**Fig. 1 Fundamental ontology in this paper.**

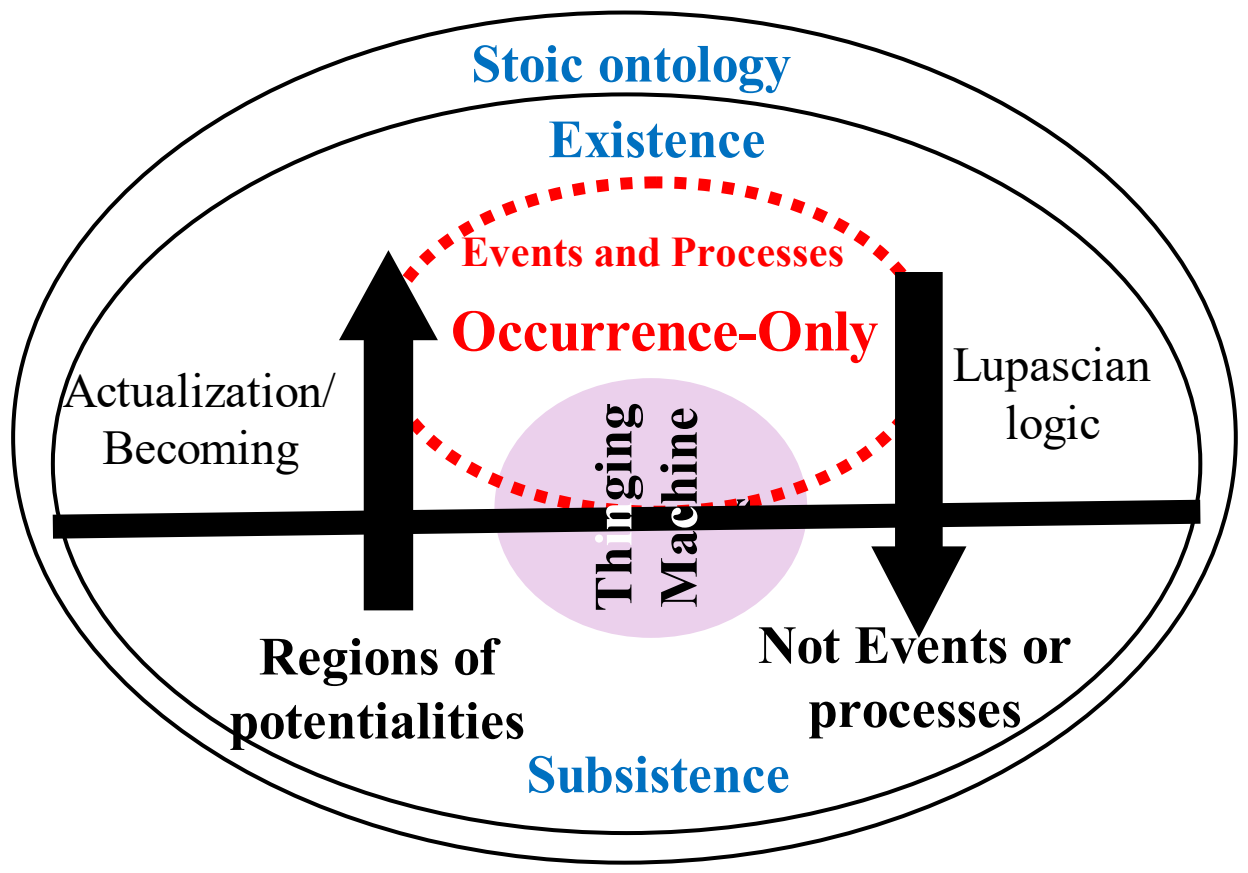

**Fig. 2 A general framework of occurrence-only modeling.**

### *A. Motivations*

To demonstrate this occurrence-only conceptual modeling, we apply it to (1) a UML/BPMN inventory system in simulation engineering and (2) an event-based system that represents medical occurrences that arise on a timeline.

An active research community in simulation engineering demonstrates the growing interest in conceptual modeling for simulation [9]. According to Wagner [10], "since a running computer simulation is a particular kind of software system, we may consider simulation engineering as a special case of *software engineering*." Modeling is an important first step in a simulation project; it is also thought to be the least understood part of simulation engineering [11]. There is a lack of standards for procedures, notation, and model qualities, and "often no information or process models are produced, but rather the modeler jumps from her mental model to its implementation in some target technology platform" [10].

In the second case study, according to Li et al. [12], in the clinical domains, temporal information elucidates the occurrence or changing status of medical events (e.g., visits, laboratory tests, procedures). Accurate profiling of clinical timelines could benefit condition trajectory tracking, adverse reaction detecting, disease risk prediction, etc. The widespread adoption of electronic health records provides great opportunities for accessing large amounts of clinical data. Due to the implicit nature of temporal expressions, often characterized by a considerable degree of under-specification, automatically constructing a timeline of clinical events is quite challenging. Modeling of temporal concepts and relationships that could support subsequent temporal reasoning is a crucial prerequisite to overcoming this hurdle [12].

Such problems in the fields of simulation and medical systems provide motivation to suggest a different approach to achieve two aims, proposing a possible solution for workers in both fields based on utilizing a new more "stakeholder friendly" conceptual modeling language and simultaneously providing an opportunity to experiment with features of such a language in a new field of application.

### *B. Main Thesis*

The adopted general philosophy in the occurrence-only approach is that all things are events [13]. For example, the life of such an "object" as man is "a historic route of events as the same enduring person from birth to death" [13]. Objects and events are things of the same kind [14] [13]. Anything that "exhibit[s] permanence and an abiding structure in nature must be explained in terms of events" [13]. According to McHenry [13],

> The expansion of the universe is an event, but so is the hurricane off the coast of California, the traffic accident outside my window, and the dance of subatomic particles in my cup of tea. So in addition to galaxies, bodies of land and sea, automobiles and cups of tea, there appear to be activities, happenings or episodes.

### C. *Example*

Entity-like events and process-like events (what Whitehead termed "actual entities" and "actual occasions," respectively) are the existing things of which the world is made up. Consider ***Socrates is walking*** [now], which involves the entity-like *Socrates* and the process-like *walking*. Contrary to the classical Aristotelian interpretation, walking is not "in" Socrates; rather, it is a persistent event. The event *Socrates* triggers the creation and processing of walk of the body Socrates. The assumption here is that Socrates is not just a body. For example, Socrates is discerning, caring, regretting, feeling, and warming, etc., which are not "in" his body, but each of them is some type of process in Socrates and is a region of potentiality.

Fig. 3 models *Socrates is walking*. We use the region (see Fig. 3) to represent *where the event occurs*. The actions *create* and *process* are two of the *five generic* actions, as we will discuss in section 2. The upper diagram (dynamic level) is the *Process* that includes the events of Socrates *existing* (create) and walking (create walking and process it). The time is assumed to be now.

The lower part of the figure (static level) provides the base for the realization of the events. The potentiality of Socrates *subsisting* refers to the potential capability of creating and processing walking. *Subsisting* and *existing* are Stoic terms that describe a view of dual *being*, as we will discuss in section 2. To simplify the event diagrams, we may replace each event with its regions. Fig. 4 (top) shows three generic events.

$E_1$: There exists Socrates.
$E_2$: Walk is generated by Socrates.
$E_3$: Waking is processed (continued).

Fig. 4 shows the behavioral model of *Socrates is walking*.

Events combine with each other to form a unity for a complex of events called *Process*. We will use the capital first letter to distinguish this *Process* from *process*, which is one of the five TM actions illustrated in this example. Romero [15] called such processes "bundles of events", "The thing 'Socrates', for instance, is a cluster of events sharing their occurrence in Greece, previous to such and such other events, including processes like 'talking with Plato', and so on" [15].

### D. *Paper Structure*

The next section provides a review and some new details of the proposed occurrence-only modeling. Section 3 presents the first case study that involves modeling an inventory system in simulation engineering. Section 4 concerns the case study modeling of clinical events in a medical information system.

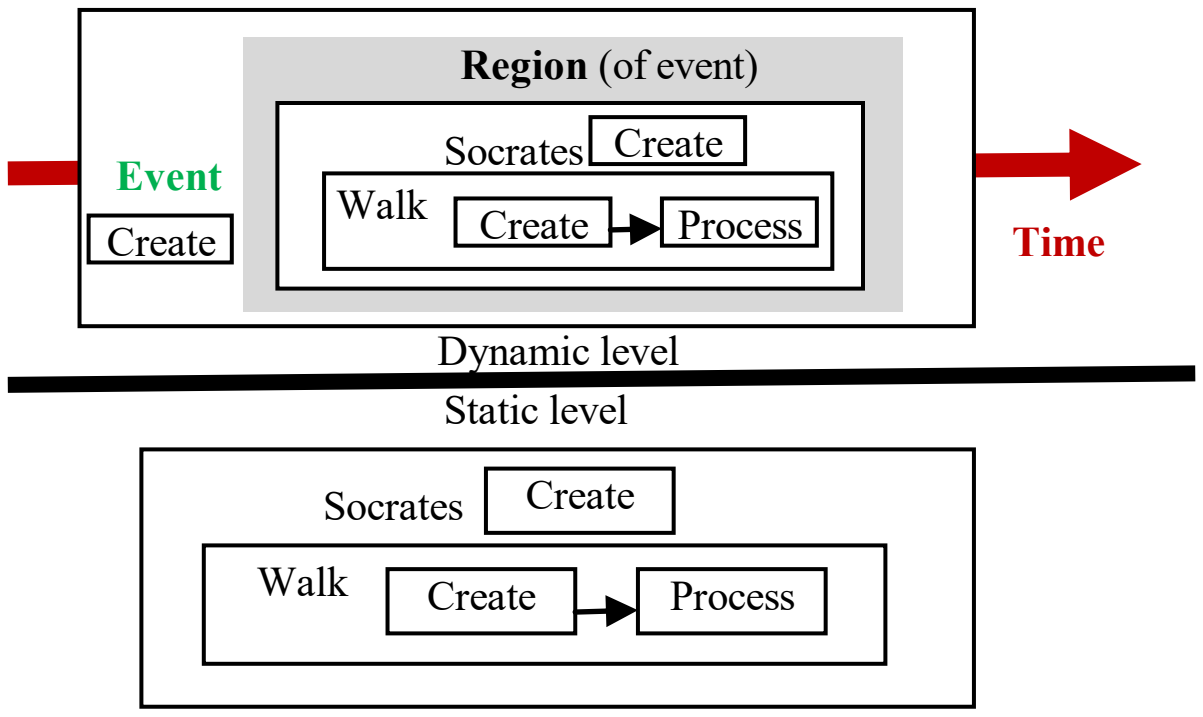

Fig. 3 Subsisting and existing Socrates walking.

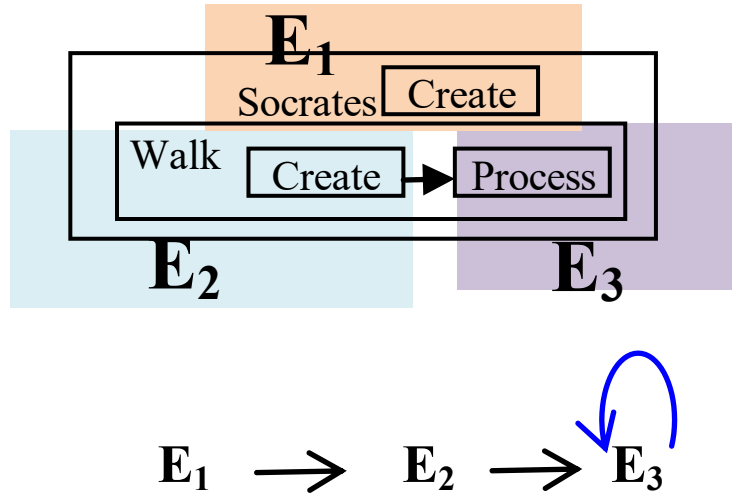

**Fig. 4 Generic events and behavior model of *Socrates is walking*.**

## II. Occurrence-Only Modeling

Occurrence-only conceptual modeling is founded on three grounds, Stoic ontology, thinging machines (TMs), and Lupascian logic. In the following, we present further details of these foundations.

### A. *Stoic Ontology*

According to Verdonck et al. [16], conceptual models lacked an adequate specification of the semantics of the terminology of the underlying models, leading to inconsistent interpretations and uses of knowledge. To provide a foundation for modeling, ontologies were introduced. Ontology would express a domain's fundamental elements and therefore would become the theoretical basis of conceptual modeling. For instance, ontological theories, such as Bunge ontology, have been used to supplement conceptual modeling languages (e.g., UML) [16]. Occurrence-only modeling is based on the Stoic ontology, which provides two levels of *being* necessary to represent reality: subsistence and existence. Stoic ontology is a materialist or, more precisely, corporealist ontology. According to such ontology, only *bodies* exist because only bodies have the capacity to act or be acted on [17]. Stoic ontology includes *bodies* that *exist* as well as entities categorized as *incorporeal* that are said to *subsist* but not to exist. These entities are nonexistent in that they are not themselves solid bodies, but they have a derivative mode of reality.

In our embracing of this ontology, existence (*what is occurring or the actual reality of being*) includes two kinds of dynamic entities: (a) enduring (extended in time) entity-like existence (e.g., electrons and subatomic particles) and (b) Process-like existence (e.g., hunting (process) and traffic jams).

**Example:** Consider the nature of software as illustrated in Fig. 5. The software is in *subsistence* while it is stored as a list of instructions. It *exists* when it is executed. In both cases, it is a thing in reality.

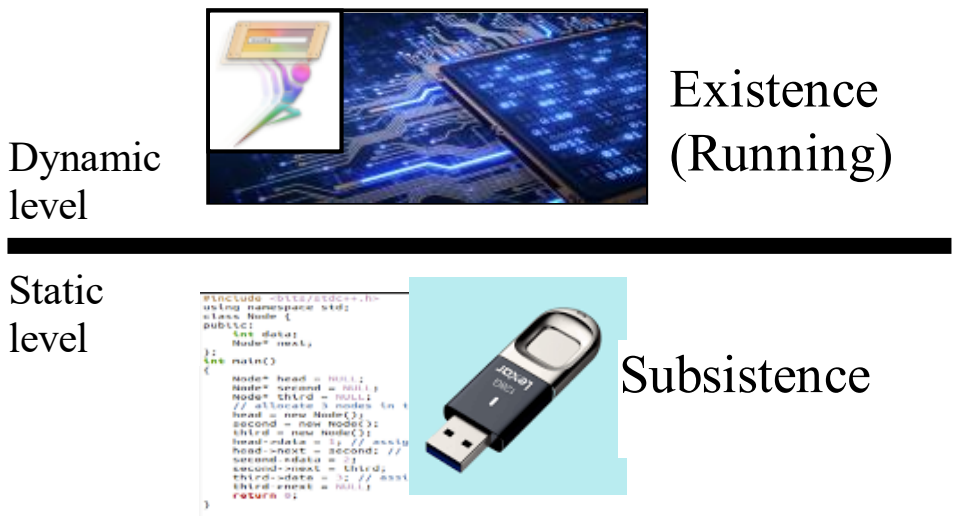

**Fig. 5 Software is in subsistence while it is not executed.**

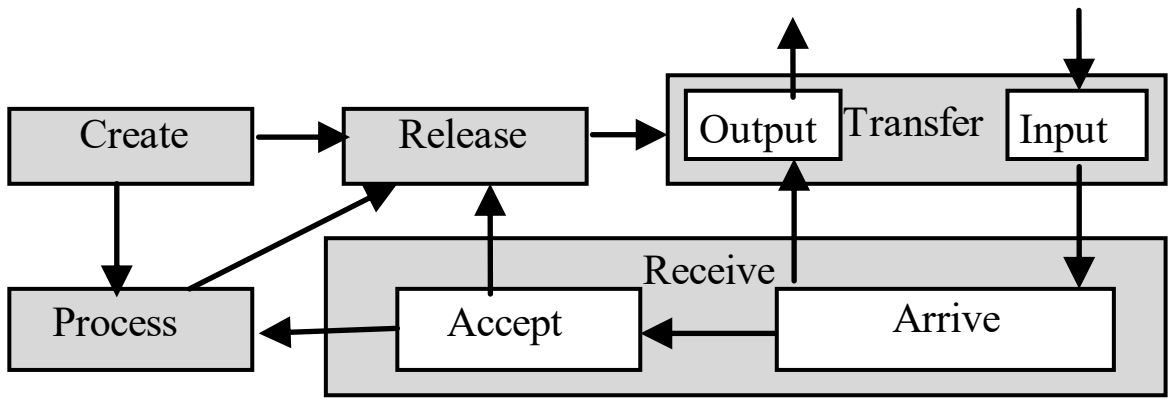

Fig. 6. Thinging machine

### B. *Thinging Machines (TM)*

In TM modeling, a *thing* is a Heideggerian notion [18] that indicates *something*. According to the Stoic doctrine, a something has a greater extension than *being*, which includes within itself the bodies and the incorporeals “entering” into the world [19]. This “entering” into the world marks “the situated-ness of the thing among other things in the world” [20].

The TM thing with this Heideggerian and Stoic underlining is called a thimac (*thi*ng/*mac*hine) because it is also conceptualized with the dual nature of a thing and machine. Such a characterization parallels the Stoic notion of a thing’s capacity to act or be acted on. However, TM comprises five actions: *create*, *process*, *release*, *transfer* and *receive* (see Fig. 6). A thimac as a thing is created, processed, released, transferred, and received. A thimac as a machine creates, processes, releases, transfers, and receives other things.

A thimac’s structure is a net of nodes. Each node has the dual structure of things and machines; therefore, these nodes are subthimacs. The thimac and its subthimacs may be connected internally and externally (outside the containing thimac) by links of flow of things. A thimac can accommodate existent, subsistent, and the other types of things that do not subsist/exist. A subsistent thing lacks a time subthimac.

The TM machine, at the static level, has the five *potential* actions: create, process, release, transfer, and receive, described as follows.

1) *Accept:* A thing enters the machine. For simplification, we assume that arriving things are *accepted* (see Fig. 6); therefore, we can combine the *arrive* and *accept* stages into the *receive* stage.

2) *Release:* A thing is ready for transfer outside the machine.

3) *Process:* A thing is changed, handled, and examined, but no new thing results.

4) *Transfer:* A thing is input into or output from a machine. The dynamic (not necessarily physical) “movement” (event) is from a previous region to a different region through a third region.

5) *Create:* A new thing (found/manifested) is realized at the dynamic level. Simultaneously, it also refers to the “existence” (at the dynamic level) of a potential thing (at the static level).

Additionally, the TM model includes a *triggering* mechanism (denoted by a dashed arrow in this article’s figures), which initiates a (non-sequential) flow from one machine to another. Moreover, each action stage may have its own memory storage (denoted by cylinder in the TM diagram) of things. A memory has its own five actions forming a *memory thimac*.

Note that for simplicity, we may omit *create* in some diagrams because the box representing the thimac implies its “beingness” (in the model). Additionally, note that the five generic actions become generic events at the dynamic level. Therefore, what we call *Process* emerges as aggregate comprising lower-level events. The resulting Process is different from the lower-level events that form it (e.g., as in chemical reactions). Structurally, as a thimac, this emergent Process has its own machine and therefore has its own behavior, i.e., a weight as a (sub)thimac is the sum of its subthimacs’ weights, and it can be created, processed, etc.

### C. *Two Thinging Machine Levels of Specification*

1) *Static (Subsistence) Model*: This model represents static things and static (potential) actions. A thing’s “being” at this level is a certain state of being, *subsistence* or a potential for “becoming,” i.e., “it is there,” inert, passive, waiting to *exist* when it couples with time. *Becoming* refers to transferring to the dynamic level to trigger the creation of an *event*. The static model is also the “inactive” state (e.g., dormant volcano). The static level is the *retreating* “world” of events, e.g., *doing something* becomes a negative event of *not doing* (a Lupascian logic term). A static thing could become an actual thing (event); however, *some* static (non-subsisting) things (e.g. square circle) could never become actual things. Accordingly, there are things that do not exist or subsist. Additionally, the static level includes all possibilities, just as a chess board exhibits all possible moves, including contradictory ones.

*2) Dynamic (Existence: occurrence only) Model*: Each event or process consists of a static subdiagram (region) that unfolds with *time*, leading to events, i.e., the realization of static things and actions. Therefore, the event is the existing being that was previously a subsisting being as a region at the static level. The Lupascian notion of a negative event refers to reverting to the static level from the dynamic level.

Stoic ontology serves to define the *being* (subsistence or existence) of things and actions in reality. The Stoics concocted the idea of a broader category of being: reality is made of things that *exist* and things that *subsist*. This idea retains the commonsensical notion that static and dynamic things are in some sense real. The notion of "modes of being" appears in various forms in classical logic, in which the notions of existence and subsistence appear [21]. Meinong [22] introduced Meinongian metaphysics and distinguished between being and existence. Using Stoic ontology, we view the dynamic model description as an *occurrence-only model of existence*. Therefore, reality includes occurrence-only things.

The static model represents the world of potentialities with atemporal subsistence. It is self-contained and in a state in which time and its related notions lose meaning. This *static universe* "contains everything there is or ever was or will be" (from [23], ignoring Post's metaphysical implications). Only a portion of this "everything" can become *occurrences*. Therefore, if we consider that the chess board includes all potential and non-potential plays, the subsisting plays are the legal plays and the existing plays are plays of the actual game. The castle that moves *nine* places (i.e., goes outside the board) is a non-subsisting play and therefore cannot occur.

At the dynamic TM level, events form among themselves an interacting nexus of occurrences that define, inform, and constitute all "actual" thimac beings. Things at the dynamic level may present object-like and Process-like occurrences. Process is another term for events and, more specifically, a net of events that forms a whole notion. For example, *release-transfer* may be considered the Process of *input*, and transfer-receive be the Process of *output*; however, *release-transfer-transfer* does not seem identified with a standalone *notion*.

The *event*, as a generic event or Process, can be provisionally defined as a fundamental happening that forms the basic building blocks of the existing world. Everything in the world, including people and things, can be constructed from events that form the essential and sole ontological elements of existence.

### *D. The Thing Side of the Thimac*

The thimac is a whole that is more than the sum of its parts (i.e., it has its own machine). Even if interiority has no subthimacs (e.g., empty safe), the thimac has some of its actions. A thing's *subsistence* means, along with its related actions, it is a potential event. An example of this subsistence is a city on a map. The city on the map can be described in terms of streets, population, connections with other cities, interaction with the environment, windiness, water resources, etc., but it is just a map with no activities. Even though it is connected with another city, there are no moving cars on the highways and no playing children in the streets. "Relations" between subsisting things are like dry river beds. Even though a dry river (e.g., release, transfer, transfer, receive) looks "permanent" in the static model, it becomes a flash *event* that may perish at any time, i.e., alternate between static and dynamic levels.

Only thimacs that embed time are realizable (exist) at the dynamic level. Therefore, for example, a "square circle" is a static thimac that cannot be injected with time to exist in the dynamic model; neither does it subsist because it is not mappable to the dynamic level. The universe of such a world is populated by things that may alternate between two levels of being: static and dynamic. This total universe is a Process (an orchestrated net of events) in which events occur and then perish or cease to be.

### *E. Lupascian Logic*

The event is different from similarly named notions currently used in the literature. Note that this approach takes the side of philosophers who conceive of physical things as extended across time (e.g., Whitehead). Objects and events are things of the same kind [24].

Therefore, instead of *doing* vs. *stop doing* (action vs negative action), we have an event, doing, that includes its region in the dynamic level vs. *stop doing*: reverting (the event's region) to static level. This method of eliminating negativity stems from philosopher Stéphane Lupasco. According to Brenner [25], every element *e* (an event, i.e., a thimac that contains a region plus time) always associates with a *non-e* (static thimac), such that the actualization of one entails the potentialization of the other and vice versa, alternatively, "without either ever disappearing completely."

With this ontological foundation of the occurrence-only modeling, the next two sections demonstrate that such a modeling approach has the expressive power to represent reasonably sized systems.

## III. MODELING AN INVENTORY SYSTEM

Wagner [10] considered a simple case of inventory management: a shop selling one product type. The customers come to the shop and place their orders. If the ordered product quantity is in stock, customers pay their order, and the ordered products are handed out to them. Otherwise, the order may still be partially fulfilled if there are still some items in stock. When the stock quantity falls below the reorder point, a replenishment order is sent to the vendor for restocking the inventory, and the ordered quantity is delivered.

Wagner [10] used a BPMN-based process design modeling approach with UML class diagrams (see Fig. 7) to develop discrete event simulations. Wagner [10] justified the use of BPMN as follows:

Using BPMN as a basis for developing a *process design modeling* approach is the best choice of a modeling language we can make, considering the alternatives, which are either not well defined or not sufficiently expressive (Italic added).

Although such an object-oriented approach is a valuable effort in applying modeling in simulation, the resultant mixed (dynamic vs. static) representation and ontological ambiguity (event vs. object) seem to produce a heterogeneous notation that distorts the purpose of the conceptual modeling as “a bridge between the developer and the user” [9] and ‘“the agreement between the simulation developer and the user about what the simulation will do” [26].

Wagner [10] is mainly concerned with *discrete event simulation*, *event process modeling* notation, and *object event* graphs. Such an *event-intensive* approach involves objects and a discrete flow of *events* that allegedly *change* the *state* of affected objects and cause follow-up *events* and a *state transition system* where *events* are transitions and the system state consists of object states and future *events*. Ontologically, this understanding of events is based on Casati and Varzi’s [27] description that Wagner [28] described as such: “The world consists of objects and events. Smiles, walks, dances, weddings, explosions, hiccups, hand-waves, arrivals and departures, births and deaths, thunder and lightning: the variety of the world seems to lie not only in the assortment of its ordinary citizens—animals and physical objects, and perhaps minds, sets, abstract particulars—but also in the sort of things that happen to or are performed by them.”

Nevertheless, Casati and Varzi [27] stated that “there is significant disagreement concerning the precise nature of such entities. (Their broad characterization as ‘things that happen’, though commonly found in dictionaries, merely shifts the burden to the task of clarifying the meaning of ‘happen’.)” Additionally, such a process-infected approach to modeling does not present or derive a clear definition of the notion of process.

The basic assertion in this paper is that using the so-called *process design* is better represented with the occurrence-only modeling. Accordingly, the resultant conceptual models settle this issue when put side by side.

### *A. Static Model*

Fig. 8 shows the basic static model of the inventory system. Basic, here, means that it is possible to enhance such a model with other details such as constraints and rules because the involved modeling language is rich in expressibility. The main stream of actions in Fig. 8 is where the customer (circle 1) creates (2) an order that flows to the shop (3) to be processed (4). Note that the order may include many data; thus, it is initially processed (the pink process box) to trigger extraction of the order *quantity*.

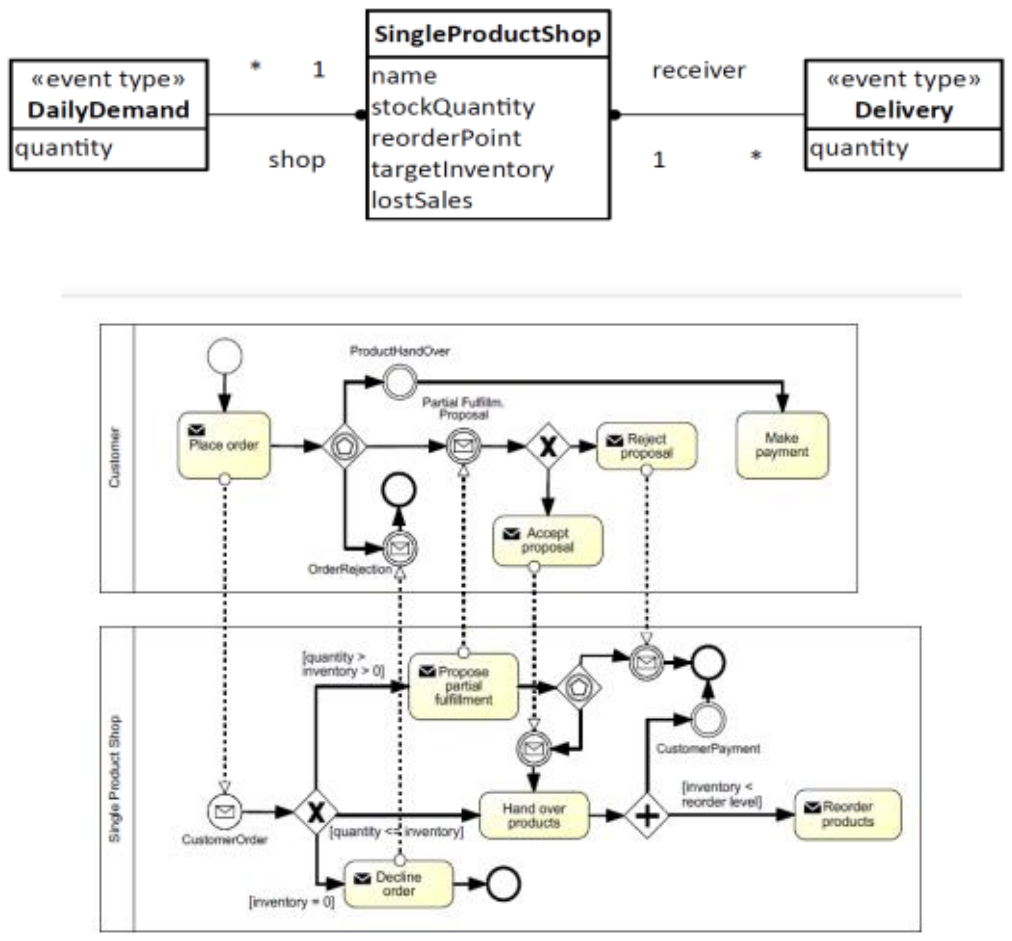

**Fig. 7 UML and BPMN diagrams used to model the inventory system (From [10]).**

The darkened boxes in the figure indicate modules in the system. The pink-shaded box in the middle of the figure is a module where main procedure is performed.

The Process (4) in the pink rectangle involves comparing the current value of the number of items in the inventory (5) with the ordered quantity. This current inventory value flows (6) to be processed (4). The Process (4) involves deciding the three following cases: Inventory = 0 (7), (customer) Quantity <= Inventory (8) and (customer) Quantity > Inventory > 0 (9).

(a) **Inventory=0 (7)**: A decline notification is created (10) and communicated to the customer (11).

(b) **(Customer) Quantity <= Inventory (8)**: This result involves two series of actions.

- An invoice is created (12) and sent to the customer (13). The customer processes it (14) to create payment (15) that is sent to the shop (16 and 17).
- The shop triggers (18) the inventory to deliver the product to the customer.

Assuming that the above two series of action are accomplished (19), the inventory sends the ordered product to the customer (20, 21, and 22). Additionally, the inventory is updated as follows.

- The ordered quantity to be delivered (the pink box) is extracted (19) and sent to the inventory (20) to be processed (21) along with the current value of the inventory to update the value (22).
- Also, the new value is processed (23) to determine whether it has reached the reordering level (24), and if it has, a reordering is created and sent to the supplier (25).
  - In case a shipment comes from the supplier (26), the current inventory value is retrieved (27) and updated (28).

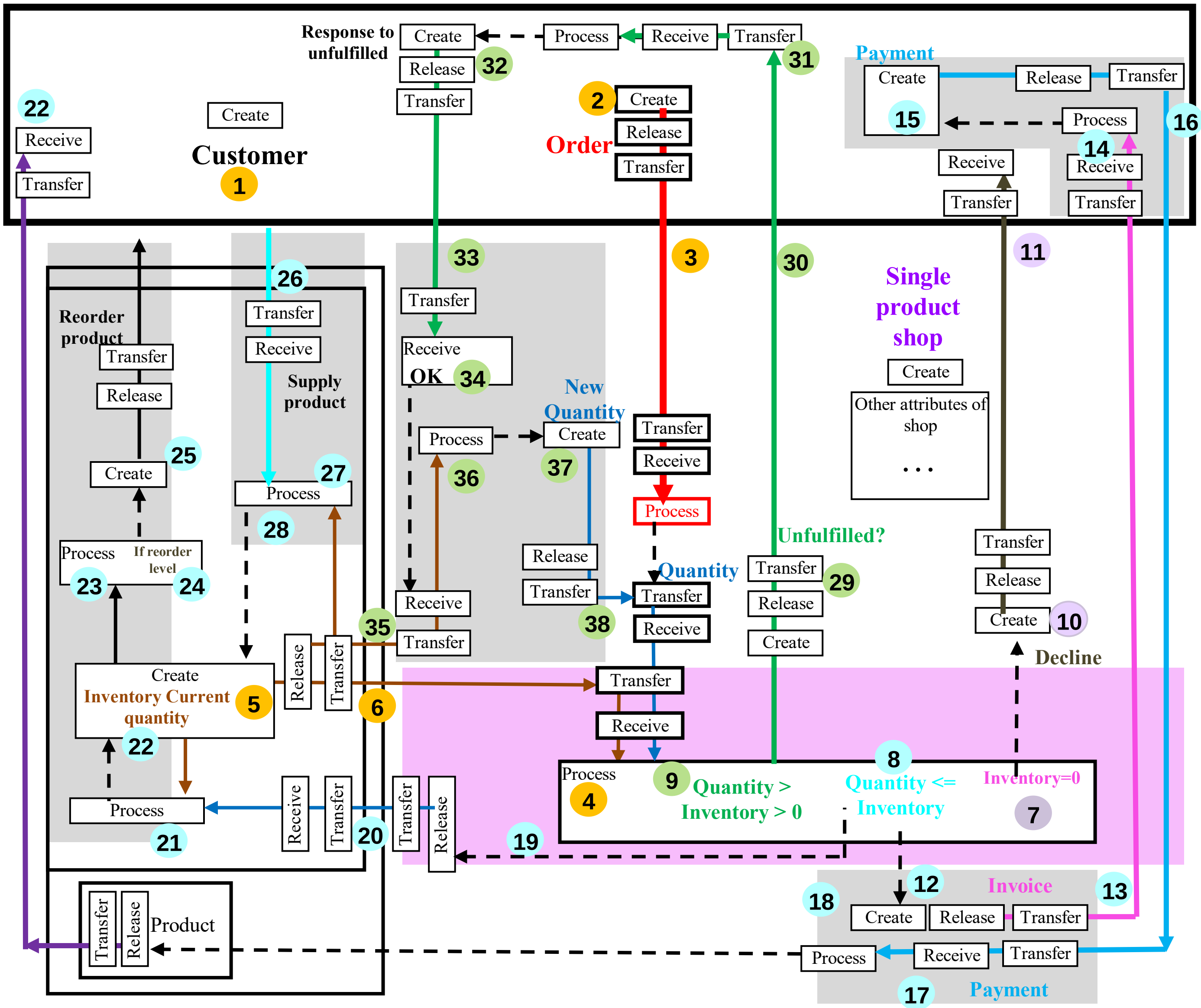

**Fig. 8 The static model of the inventory system.**

(c) **(Customer) Quantity > Inventory**: A notification is created (29) and sent to the customer (30). The customer processes (31) the notification and creates a response (32) that flows to the shop (33). Assuming that the partial fulfillment is okay (34, the current value of inventory is retrieved (35), processed (36), and inserted as a new ordered quantity (37). Hence, the customer order is processed (with its new value) again (38) where ordered quantity is equal to the inventory value.

Fig. 8 is an engineering diagram that will be realized as a tangible Process. It looks to be a complex diagram; however, complexity is a relative term. When two representations involve the same level of abstraction, we can say that one of them is more complex than the other. UML is known for its complexity because it involves 14 models, each with different notations.

There are no generally accepted semantics of these concepts as conceptual modeling elements [29]. On the other hand, the apparent complexity of Fig. 8 appears as the result of repeatedly using the five generic actions *create process*, *release*, *transfer*, and *receive*, which give the model a uniformity that is rarely found in systems.

Fig. 8 can be simplified by assuming that the arrow direction indicates the direction of flow; thus, the transfer, release, and receive actions can be eliminated, resulting in Fig. 9. Note that the original diagram is still the base for the design phase, just as a complex electric circuit may be simplified by using such a technique as combining series and parallel resistor within the context of the larger circuit. Furthermore, this simplified diagram can be further simplified, e.g., eliminating *create* and *process*.

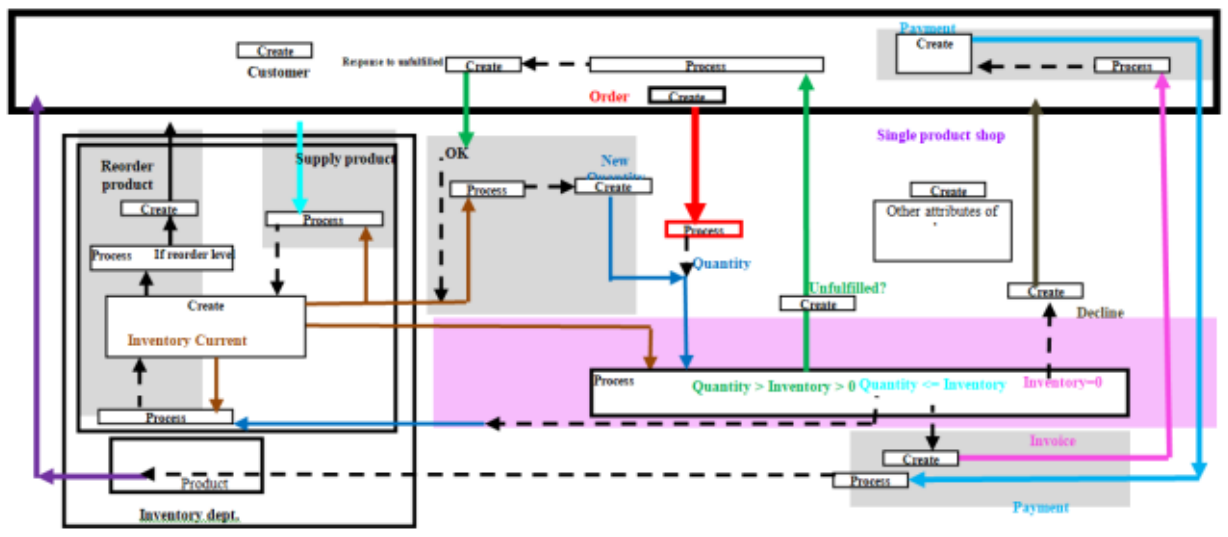

Fig. 9 Simplified static model of the inventory system.

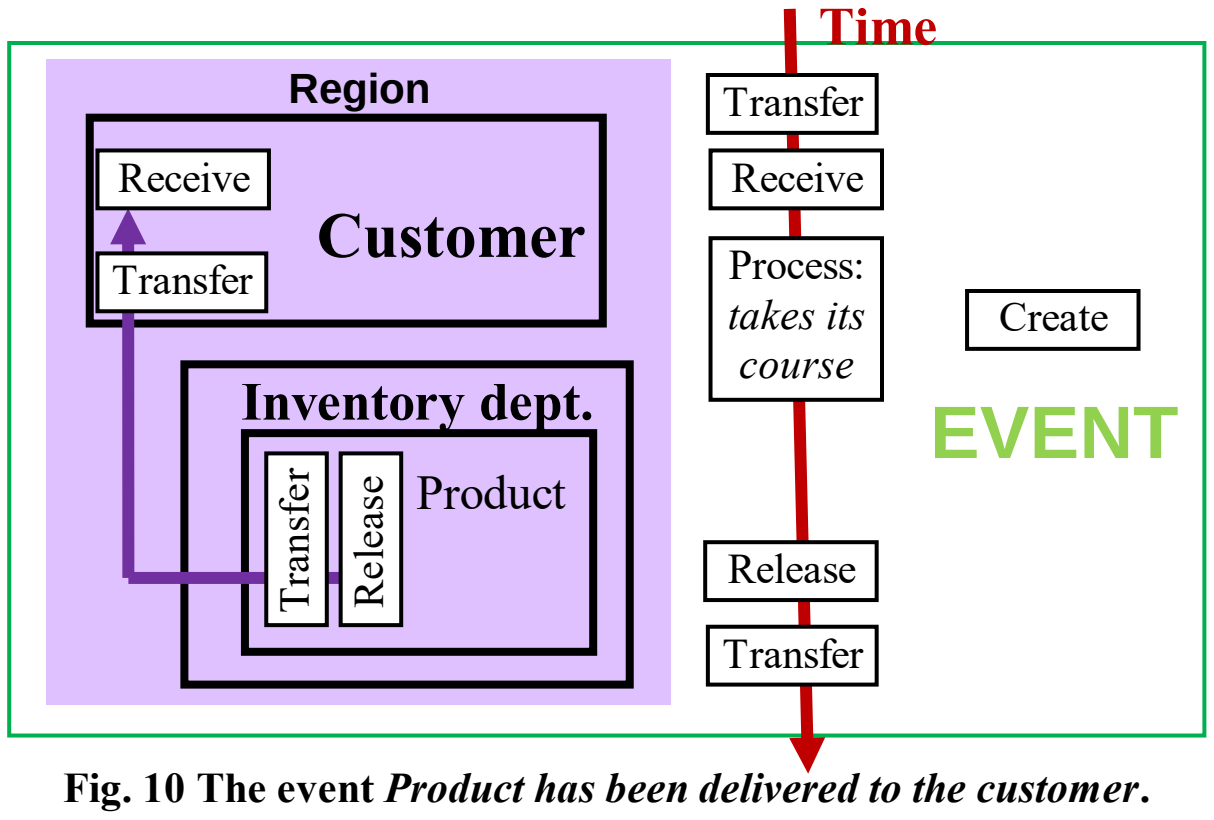

Fig. 10 The event *Product has been delivered to the customer*.

### *B. Dynamic and Behavior Models*

An event is a subdiagram of the static model (called region of event) injected with time. Fig. 10 shows the description of the event *Product has been delivered to the customer*.

For simplification sake, we will represent an event by its region. Accordingly, we identify the following events that are shown in Fig. 11.

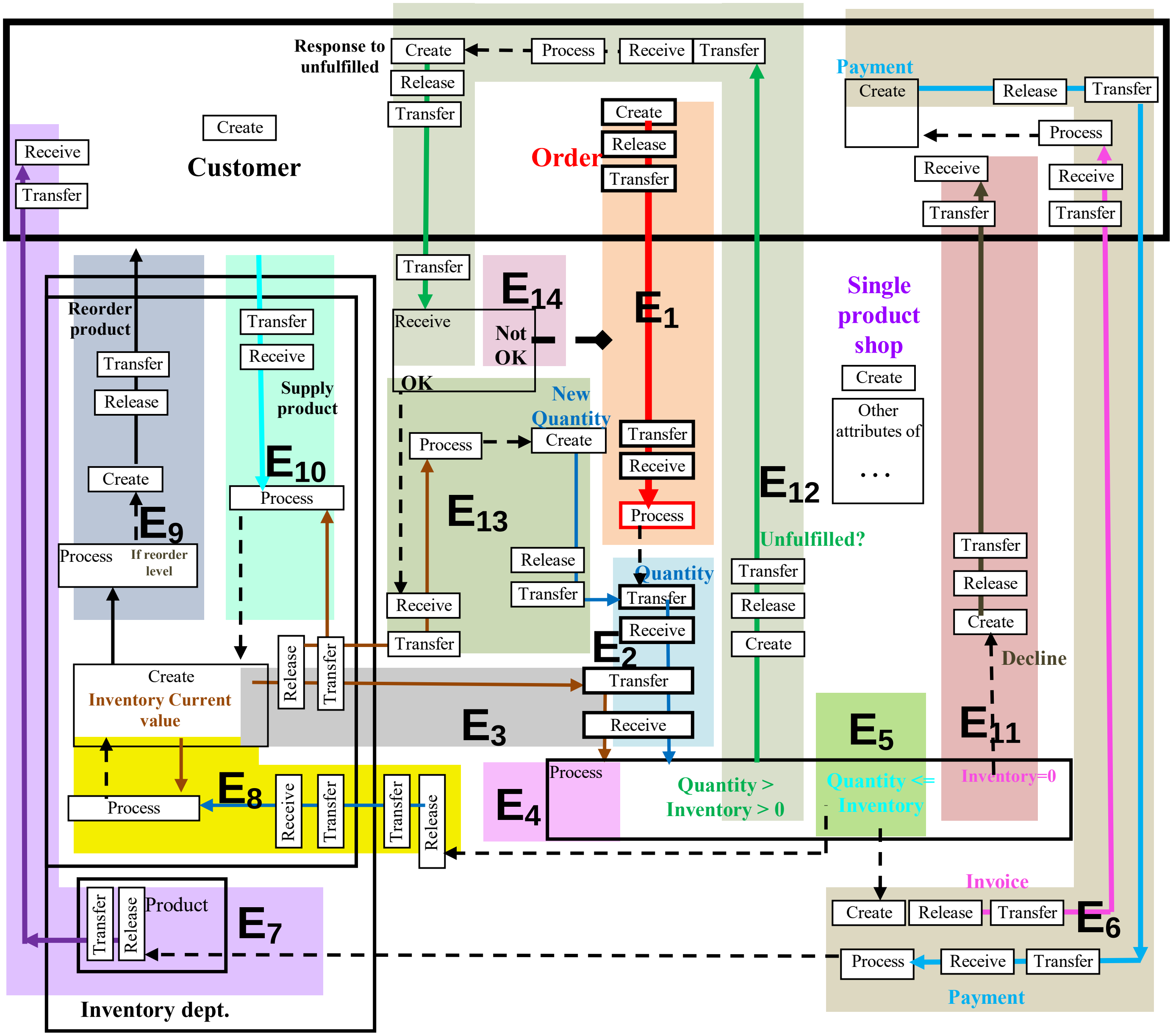

Fig. 11 The dynamic model of the inventory system.

$E_1$: The customer creates an order that flows to the shop to be processed.
$E_2$: The ordered quantity is extracted from the order.
$E_3$: The current inventory value is retrieved.
$E_4$: The ordered quantity is compared with the inventory.
$E_5$: The result of comparison is Quantity <= Inventory.
$E_6$: Invoice is sent and a payment is received.
$E_7$: Product has been delivered to the customer.
$E_8$: Inventory Current value has been updated.
$E_9$: Reordering level has been reached; hence, a supply order has been sent to the supplier.
$E_{10}$: Ordered product from the supplier is received and the inventory value is updated.
$E_{11}$: The result of comparison is Inventory = 0; hence, a decline notification is sent to the customer.
$E_{12}$: The result is Quantity > Inventory > 0; hence, a confirmation of partial fulfillment is sent to the customer.
$E_{13}$: The customer accepts partial fulfilment.
$E_{14}$: The customer does not accept partial fulfilment; hence, the order is cancelled.

Fig. 12 shows the behavior model of the inventory system. Note how the customer order is cancelled in case of the customer's refusal of a partial fulfillment ($R_1$ – This means reverting to region 1; that is, the order no longer exists). This cancelation is represented by a diamond-tail arrow from $E_{14}$ to $E_1$. This means, according to Lupascian logic, "not $E_1$," which means returning to subsistence in Stoic ontology. Semantically, this indicates that the customer order does not exist anymore.

### *C. Queuing as a Process*

Consider the *Process* where it is required to install a queue of orders waiting to be processed to extract the order quantity (red process box in Fig. 8). Fig. 13 shows how to install such a queue just before this process. We only show the dynamic model to save space since the static model can be extracted from the dynamic model.

- In the figure, $E_1$ and $E_2$ are the two events of receiving the orders, inputting them into the queue Q, and making Q *not empty* ($E_3$). This procedure continues filling the Q without limit (assumption). An empty Q is an initial condition.

- If the Q is *not empty* ($E_3$) and the Process (Red box) is *not busy* ($E_4$) then an order is retrieved from Q in Process ($E_5$) and sent to the Process (red box). If $E_5$ leaves the Q empty ($E_6$), then the Q indicator is set to *empty* ($E_7$).

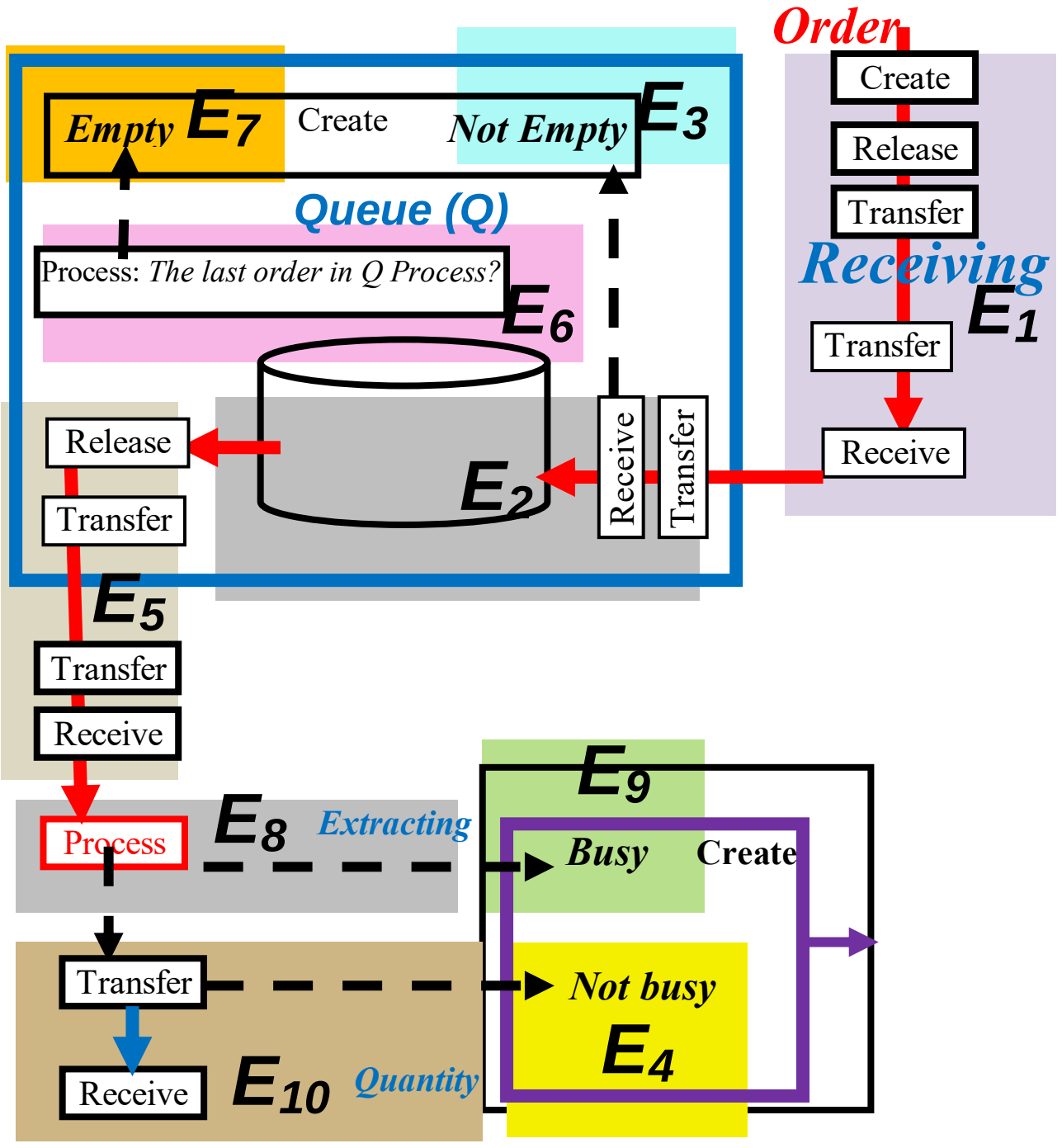

**Fig. 13 The dynamic model of the queue system.**

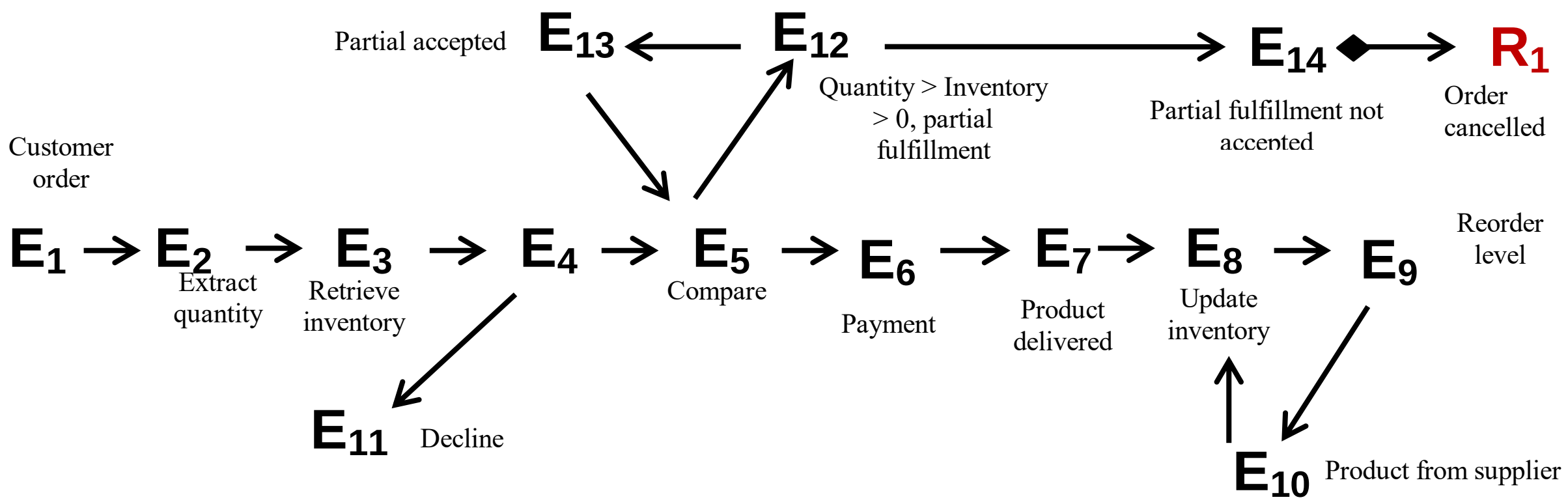

**Fig. 12 The behavior model of the inventory system.**

- The action process (red box *process* in Fig. 13) is initially *not busy*. When the Process is activated ($E_8$), its indicator is set to *busy* ($E_9$). When the Process (red box) finishes ($E_{10}$), its indicator is set to *not busy* ($E_4$).

Fig. 14 shows the behavioral model of this queuing Process.

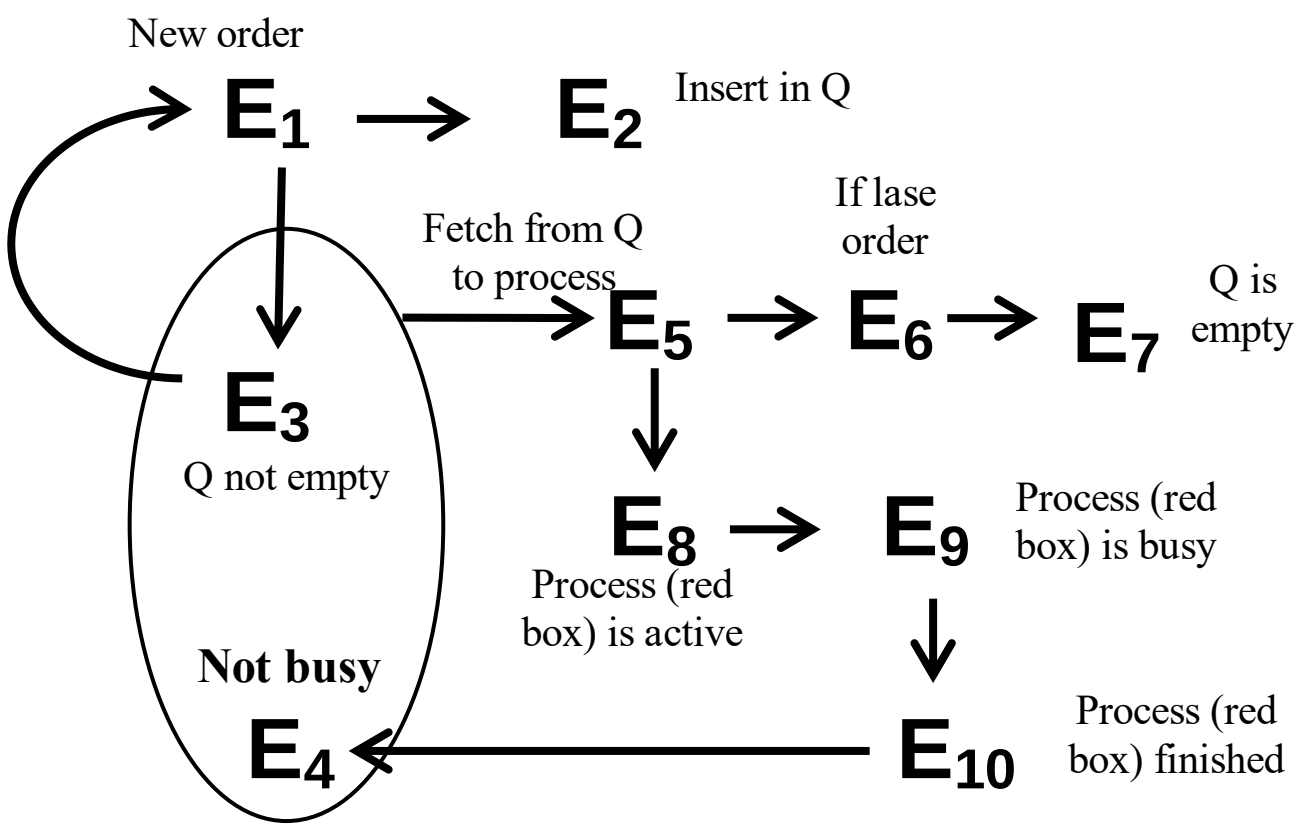

Fig. 14 The behavior model of the queue system.

## IV. Modeling Medical System

According to Li et al. [12], "Time is an important and pervasive concept of the real world. Li et al. [12] developed a time event ontology with "a rich set of classes and properties (object, data, and annotation)" that can formally represent and reason both structured and unstructured temporal information. They used the following:

- Concept primitives: clinical events "(anything that is relevant to the patient's clinical timeline) and temporal expressions and 'enriched' temporal relations."
- Real electronic health record data that faithfully represent more than 95% of the temporal expression, according to Li et al. [12].
- There are six types of events: test, problem, treatment, clinical_dept, evidential, and occurrence [30].

The results applied to a set of frequently asked time-related queries that show a strong capability of reasoning complex temporal relations.

Li et al. [12] introduced a class event to represent time-oriented medical events, which include any sort of "occurrences, states, procedures or situations that occurs on a timeline." Several subclasses are designed to cover the common clinical events (e.g., clinical intervention, diagnosis, test).

As an example, the following events report was initially manually annotated and then loaded into the Reasoner for inference. In the report, the words in red italic are manually annotated as events.

> A 35-year-old man was admitted to hospital with periorbital swelling, redness, and pain on May 24, 2014. Then he was diagnosed with periorbital cellulitis. He was treated with intravenous (IV) clindamycin, and with IV ciprofloxacin, which reduced the orbital redness and swelling. However, on the second day following antibiotic treatment, he developed nausea and right upper quadrant (RUQ) abdominal pain, his liver function tests (LFTs) began to increase. A diagnosis of idiosyncratic drug-induced liver injury (DILI) was made. [12]

### A. *Static Model*

Fig. 15 shows the corresponding static SCM model. First, the patient is admitted to the hospital (number 1) to be processed (2) and to record the patient's data (3). Note that to add some structure to the hospital, reception (4) and emergency (5) are added.

For example, this would give justification for executing to consecutive diagnoses at the beginning. The red arrow represents the movement of the patient through different stages of the medical processes. The first process (6) triggers the creation of initial diagnoses (7). Then, a diagnostic process triggered a medical description (8 and 9). In (10), a process created a prescription (11), thus triggering the delivery (e.g., from pharmacy) of medicine (12) to the patient (13). Accordingly, "Orbital redness and swelling" is reduced (14). This is followed by another process of diagnoses (15) to discover that "nausea and right upper quadrant (RUQ) abdominal pain, his liver function tests (LFTs) began to increase" (16). As a treatment (17), a prescription is written (18).

Note that the patient is a thing (red arrow) that goes through all of these processes, and, at certain stages, the relevant data of that point appear. For example, at (13 and 14), the patient is "expanded" to indicate the execution of medicine prescribed in (12) and the appearance of new patients' symptoms.

### B. *Dynamic Model*

The following events are selected (Fig. 16).

$E_1$: The *patient* is admitted in the hospital and necessary data are recorded.

$E_2$: *Initial* diagnoses: "Periorbital swelling, redness, and pain"

$E_3$; Patient is examined and diagnosis is "periorbital cellulitis."

$E_4$: A prescription is written.

$E_5$: Medicine is given to the patient.

$E_6$; Orbital redness and swelling are reduced.

$E_7$: *"Nausea and right upper quadrant (RUQ) abdominal pain, his liver function tests (LFTs) began to increase"* began to increase.

$E_8$: Prescribed treatment *"idiosyncratic drug-induced liver injury (DILI)."*

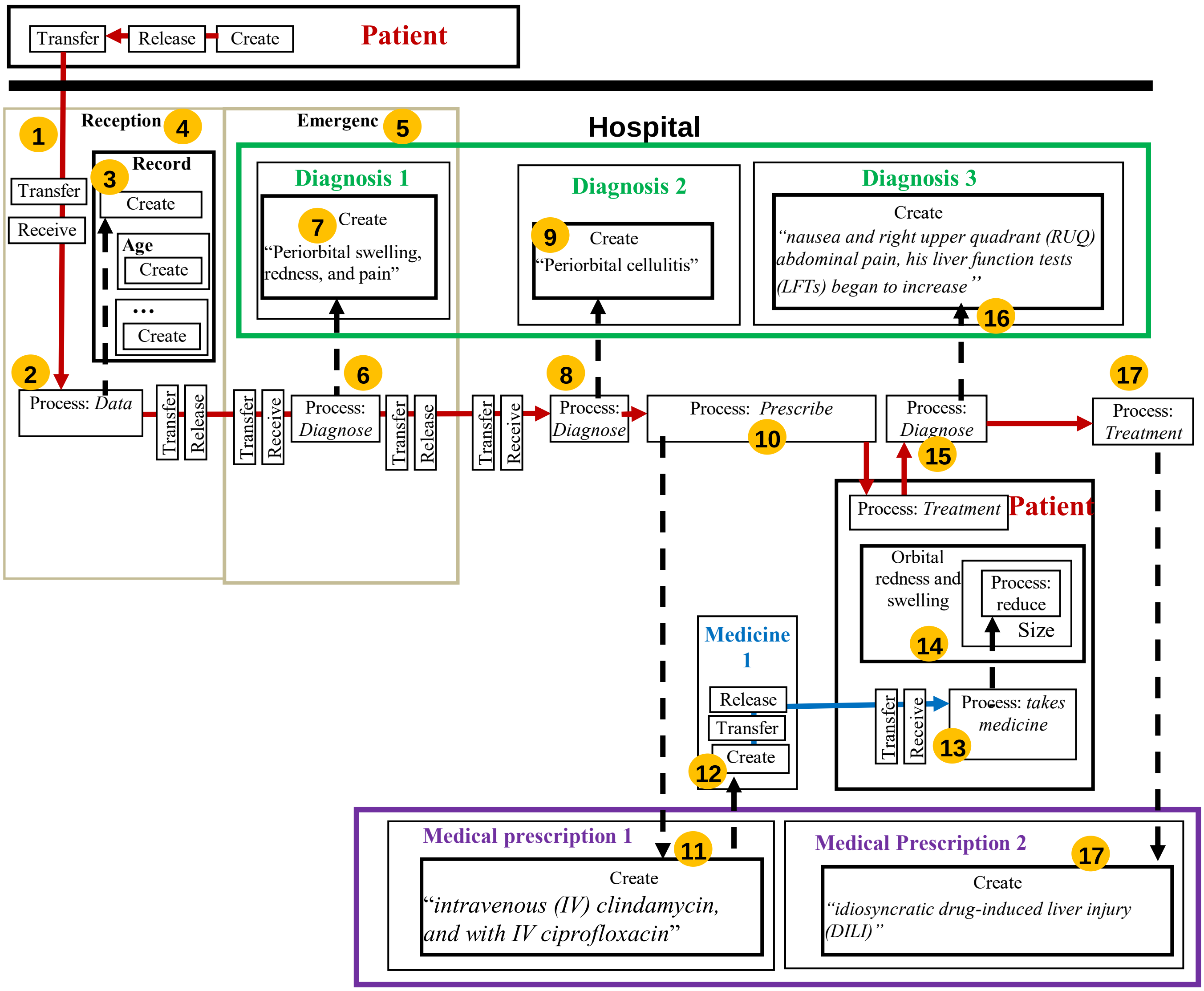

Fig. 15 The static model.

### C. *Analysis*

It is not difficult to see how this model should be generalized to become the base of a software system for any patient instead of the specific man mentioned in the events report given by Li et al. [12]. For example, diagnoses may be included in one file (e.g., UML class) instead of just three diagnoses marked in green in the model. Similarly, prescriptions are stored together (purple box includes medical prescriptions 1 and 2).

Li et al. [12] also gave sample queries that can be applied for events given in the events report. These and others can be incorporated into the SCM, including the following queries given by Li et al. [12].

*Query 1*: When was the patient admitted to the hospital? (**Answer is in E$_1$**)

*Query 2*: What is the temporal relation between "admitted to hospital" and "liver function tests (LFTs) began to increase"? (**E1 and E7**)

*Query 3*: Does "ciprofloxacin" treatment start before "diagnosis of Does "ciprofloxacin" treatment start before "diagnosis of idiosyncratic drug-induced liver injury (DILI)"? drug-induced liver injury (DILI)"? (**E4 and E8**)

*Query 1*: What events happened before "diagnosis of idiosyncratic drug-induced liver injury (DILI)"? (**E1 to E8**)

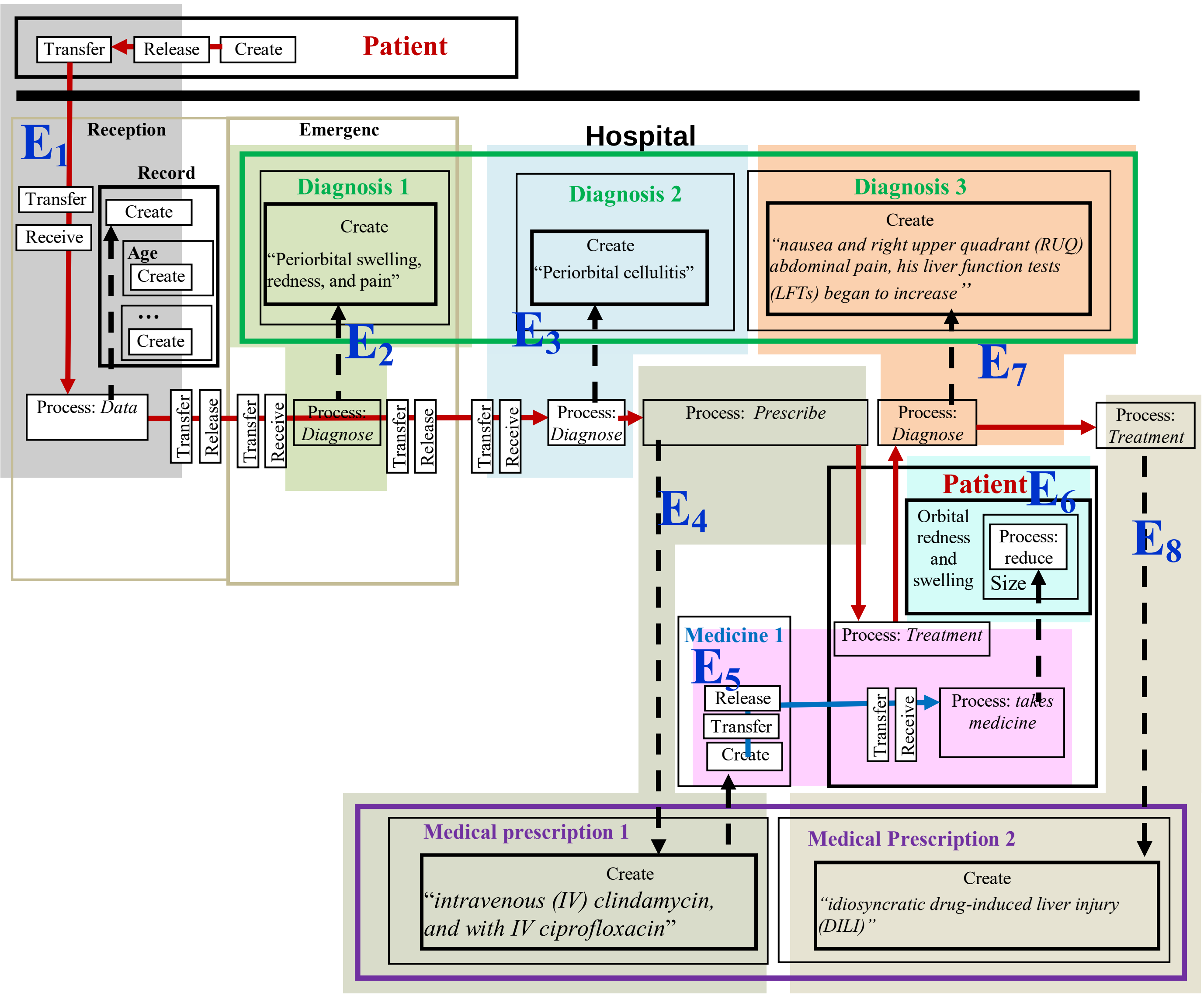

Fig. 16 Dynamic model

## V. Conclusion

The occurrence-only paradigm presented in this paper refers to conceptual modeling solely based on events and processes. This model has five generic events with high-level events formed from these generic events. Some of these high-level events are processes when the complex of events has semantically whole. For example, the inventory control system discussed in section 3 can be called a process, whereas an arbitrary subdiagram of it may not form a "whole" with associated events that may not be qualified with a specific name.

The occurrence-only modeling can be categorized as an anti-thesis of the currently dominant object-oriented conceptual modeling (individual-based modeling with a commitment to message passing, encapsulation, inheritance, etc.). Although the basic idea of incorporating events and processes in modeling has been utilized by many researchers, the occurrence-only approach is probably the first attempt to build a "top-down" modeling ontology and language based on these two notions as first-class citizens. Hence, no claim of completeness or correctness can be applied for such a venture.

Accordingly, details and scrutiny of some parts may uncover ambiguity and errors at different portions of the modeling technique. Hopefully, pursuing further refinements though modeling applications in different domains would uncover these ambiguities and errors.

In the ontology part, the subsistence notion needs further scrutiny, especially the reasons for its rejection by reputable philosophers. The thing/machine concept requires further refinement such as a situation that cannot expressed by the five-action machine.